\documentclass[pre,reprint,twoside,graphicx]{revtex4-1}

\usepackage{graphicx,epstopdf,amsfonts,amsmath,amssymb,array,mathrsfs,mathtools,amsthm,enumerate}

\usepackage[percent]{overpic}

\usepackage[textwidth=8em,textsize=small]{todonotes}

\usepackage{titlesec}

\usepackage{fancyhdr}

\def\be{\begin{equation}}
\def\ee{\end{equation}}

\def\cvp{\raise 2pt\hbox{,}}  \def\cvd{\raise 2pt\hbox{.}} 

\newcommand{\textbfsf}[1]{{\bf \textsf{#1}}}

\titleformat*{\section}{\bfseries\sffamily}
\titleformat*{\subsection}{\bfseries\sffamily}

\def\cov{{\mathrm{cov}}}
\def\BM{{\mathsf B}}
\def\X{{\mathsf X}}
\def\Y{{\mathsf Y}}
\def\prob{{\mathbf{P}}}
\def\bcov{{\mathcal{B}}}
\def\x{{\mathsf x}} 
\def\y{{\mathsf y}}

\begin{document}


\title{\Large\textbfsf{Brownian forgery of statistical dependences}} 

\author{Vincent Wens}
\email[]{vwens@ulb.ac.be}
\affiliation{\textsl{Laboratoire de Cartographie fonctionnelle du Cerveau, UNI -- ULB Neurosciences Institute, Universit\'e libre de Bruxelles (ULB) \& Magnetoencephalography Unit, Department of Functional Neuroimaging, Service of Nuclear Medicine, CUB -- H\^opital Erasme, Brussels, 
Belgium.}}


\begin{abstract}

The balance held by Brownian motion between temporal regularity and randomness is embodied in a remarkable way by Levy's forgery of continuous functions.
Here we describe how this property can be extended to forge arbitrary dependences between two statistical systems, and then establish a new Brownian independence test based on fluctuating random paths. 
We also argue that this result allows revisiting the theory of Brownian covariance from a physical perspective and opens the possibility of engineering nonlinear correlation measures from more general functional integrals.

\noindent{\sf \begin{center} This article is published in Front.~Appl.~Math.~Stat.~4:19.~DOI: 10.3389/fams.2018.00019. \end{center}}

\end{abstract}

\maketitle

%

\pagestyle{fancy}
\fancyhead{}
\fancyhead[RO]{\footnotesize \sf Wens V (2018). Front.~Appl.~Math.~Stat.~4:19} 
\fancyhead[LE]{\footnotesize \sf Brownian forgery of statistical dependences}
\renewcommand{\headrulewidth}{0pt}
	
\renewcommand\theequation{\thesection.\arabic{equation}}

\section{INTRODUCTION AND OVERVIEW}
\setcounter{equation}{0}

The modern theory of Brownian motion provides an exceptionally successful example of how physical models can have far-reaching consequences beyond their initial field of development. 
Since its introduction as a model of particle diffusion, Brownian motion has indeed enabled the description of a variety of phenomena in cell biology, neuroscience, engineering, and finance \cite{Sethna2006book}. 
Its mathematical formulation, based on the Wiener measure, also represents a fundamental prototype of continuous-time stochastic process and serves as powerful tool in probability and statistics \cite{Freedman1983book,GikhmanSkorokhod1969book}. 
Following a similar vein, we develop in this note a new way of applying Brownian motion to the characterization of statistical independence.

Our connection between Brownian motion and independence is motivated by recent developments in statistics, more specifically the unexpected coincidence of two different-looking dependence measures: distance covariance, which characterizes independence fully thanks to its built-in sensitivity to all possible relationships between two random variables \cite{Szekely2007AnnStat}, and Brownian covariance, a version of covariance that involves nonlinearities randomly generated by a Brownian process \cite{Szekely2009AnnAppStat}. 
Their equivalence provides a realization of the aforementioned connection, albeit in a somewhat indirect way that conceals its naturalness.
Our goal is to explicit how Brownian motion can reveal statistical independence by relying directly on the geometry of its sample paths. 

The \emph{brute force} method to establish the dependence or independence of two real-valued random variables $\X$ and $\Y$ consists in examining all potential relations between them. 
More formally, it is sufficient to measure the covariances $\cov[f(\X),g(\Y)]$ associated with  transformations $f, g$ that are bounded and continuous (see, e.g., Theorem 10.1 in Ref.~\cite{JacodProtter2003book}).
The question pursued here is whether using sample paths of Brownian motion in place of bounded continuous functions also allows to characterize independence, and we shall demonstrate that the answer is yes. 
In a nutshell, the statistical fluctuations of Brownian paths $\BM$, $\BM'$ enable the \emph{stochastic} covariance index $\cov[\BM(\X),\BM'(\Y)]$ to probe arbitrary dependences between the random variables $\X$ and $\Y$.

Our strategy to realize this idea consists in establishing that, given any pair $f,g$ of bounded continuous functions and any level of accuracy, the covariance $\cov[f(\X),g(\Y)]$ can be approximated \emph{generically} by $\cov[\BM(\X),\BM'(\Y)]$. 
Crucially, the notion of genericity used here refers to the fact that the probability of picking paths $\BM$, $\BM'$ fulfilling this approximation is nonzero, which ensures that an appropriate selection of stochastic covariance can be achieved by finite sampling of Brownian motion.
This core result of the paper will be referred to as the \emph{forgery of statistical dependences}, in analogy with Levy's classical forgery theorem \cite{Freedman1983book}.

Actually Levy's remarkable theorem, which states that any continuous function can be approximated on a finite interval by generic Brownian paths, provides an obvious starting point of our analysis.
Indeed, it stands to reason that if the paths $\BM$ and $\BM'$ approach the functions $f$ and $g$, respectively, then $\cov[\BM(\X),\BM'(\Y)]$ should approach $\cov[f(\X),g(\Y)]$ as well.
A technical difficulty, however, lies with the restriction to finite intervals since the random variables $\X$ and $\Y$ may be unbounded. 
Although it turns out that intervals can not be prolonged as such without ruining genericity, we shall describe first a suitable extension of Levy's forgery that holds on infinite domains.
Our forgery of statistical dependences will then follow. 

From a practical standpoint, using Brownian motion to establish independence turns out to be advantageous.
Indeed, exploring all bounded continuous transformations exhaustively is realistically impossible. (This practical difficulty motivates the use of reproducing kernel Hilbert spaces, see, e.g., Ref.~\cite{Gretton2010JMLR} for a review.)
Generating all possible realizations of Brownian motion obviously poses the same problem, but this unwieldy task can be bypassed by averaging directly over sample paths.
In this way, and quite amazingly, the measurement of an uncountable infinity of covariance indices can be replaced by a \emph{single} functional integral.
We shall discuss how this idea leads back to the concept of Brownian covariance and how the forgery of statistical dependences clarifies the way it does characterize independence, without reference to the equivalence with distance covariance.
Brownian covariance represents a very promising tool for modern data analysis \cite{Newton2009AnnAppStat, Szekely2010AnnAppStat} but appears to be still scarcely used in applications (with seminal exceptions for nonlinear time series \cite{Zhou2012JTSA} or brain connectomics \cite{Geerligs2016NI}). Our approach based on random paths is both physically grounded and mathematically rigorous, so we believe that it may help further disseminate this method and establish it as a standard tool of statistics.

\section{MAIN RESULTS}\label{mainresults}
Here we motivate and describe our main results, with sufficient precision to provide a self-contained presentation of the ideas introduced above while avoiding technical details, which are then developed in the dedicated Section \ref{mathematicalconsiderations}. 
We also use here assumptions that are slightly stronger than is necessary, and some generalizations are relegated to \ref{strongforgery}.

\subsection{Extension of Levy's forgery}\label{extendedforgery}

Imagine recording the movement of a free Brownian particle in a very large number of trials. In essence, Levy's forgery ensures that one of these traces will follow closely a predefined test trajectory, at least for some time. 
To formulate this more precisely, let us focus for definiteness on \emph{standard} Brownian motion $\BM$, whose initial value is set to $\BM(0)=0$ and variance at time $t$ is normalized to $\langle\BM(t)^2\rangle=|t|$.
We fix a real-valued continuous function $f$ with $f(0)=0$ (the test trajectory) and consider the \emph{uniform} approximation event that a Brownian path $\BM$ fits $f$ tightly up to a constant distance $\delta>0$ on the time interval $[-T, T]$, 
\be\label{unifapprox} \mathscr U_{f,\delta,T} = \big\{  |\BM(t)-f(t)| <\delta \, , \,  \forall\, |t|\leq T \big\}\, . \ee
Levy's forgery theorem states that this event is generic, i.e., it occurs with nonzero probability $\prob(\mathscr U_{f,\delta,T})>0$ (see Chapter 1, Theorem 38 in Ref.~\cite{Freedman1983book}).
This result requires both the randomness and the continuity of Brownian motion. 
Neither deterministic processes nor white noises satisfy this property.

%
In all trials though, the particle will eventually drift away to infinity and thus deviate from any bounded test trajectory.
Indeed, let us further assume that the function $f$ is bounded and examine what happens when $T\rightarrow\infty$.
If the limit event $\mathscr U_{f,\delta,\infty}=\bigcap_{T>0}\mathscr U_{f,\delta,T}$ occurs, the path $\BM$ must be bounded too since the particle is forever trapped in a finite-size neighborhood of the test trajectory.
%
%
However Brownian motion is almost surely (a.s.)~unbounded at long times \cite{Freedman1983book}, so that
$\prob(\mathscr U_{f,\delta,\infty})=0$.
%
%
Hence Levy's forgery theorem does not work on infinite time domains.

To accommodate this asymptotic behavior, we should thus allow the particle to diverge from the test trajectory, at least in a controlled way.
Let us recall that the escape to infinity is a.s.~slower for Brownian motion than for any movement at constant velocity (which is one way to state the law of large numbers \cite{Freedman1983book}).
This suggests adjoining to event \eqref{unifapprox} the loose approximation event
\be\label{linearapprox} \mathscr E_{f,v,T} = \big\{  |\BM(t)-f(t)| <v |t| \, , \,  \forall\, |t|\geq T \big\}\ee
whereby the particle is confined to a neighborhood of the test trajectory that \emph{expands} at finite speed $v>0$. 

\medskip\noindent{\textbfsf{Asymptotic forgery theorem.}} {\it Let $f$ be bounded and continuous, and $v,T>0$. Then $\prob(\mathscr E_{f,v,T})>0$.}

\medskip\noindent An elegant, albeit slightly abstract, proof rests on a short/long time duality between the classes of events \eqref{unifapprox} and \eqref{linearapprox}, which maps Levy's forgery and this asymptotic version onto each other (see Section \ref{dualityproof}).
For a more concrete approach, let us focus on the large $T$ limit that will be used to study statistical dependences.
Since the path $\BM(t)$ and the neighborhood size $v|t|$ both diverge, the bounded term $f(t)$ can be neglected in Eq.~\eqref{linearapprox} and the event $\mathscr E_{f,v,T}$ thus merely requires not to outrun deterministic particles moving at speed $v$.
The asymptotic forgery thus reduces to the law of large numbers, which ensures that $\prob(\mathscr E_{f,v,T})$ is close to one for all $T\gg1$.
This probability decreases continuously as $T$ is lowered [since the defining condition in Eq.~\eqref{linearapprox} becomes stricter] but does not drop to zero until $T=0$ is reached.
This line of reasoning can be completed and also generalized to allow slower expansions (see \ref{strongforgery}). 

\begin{figure}
\begin{overpic}[width=8.5cm]{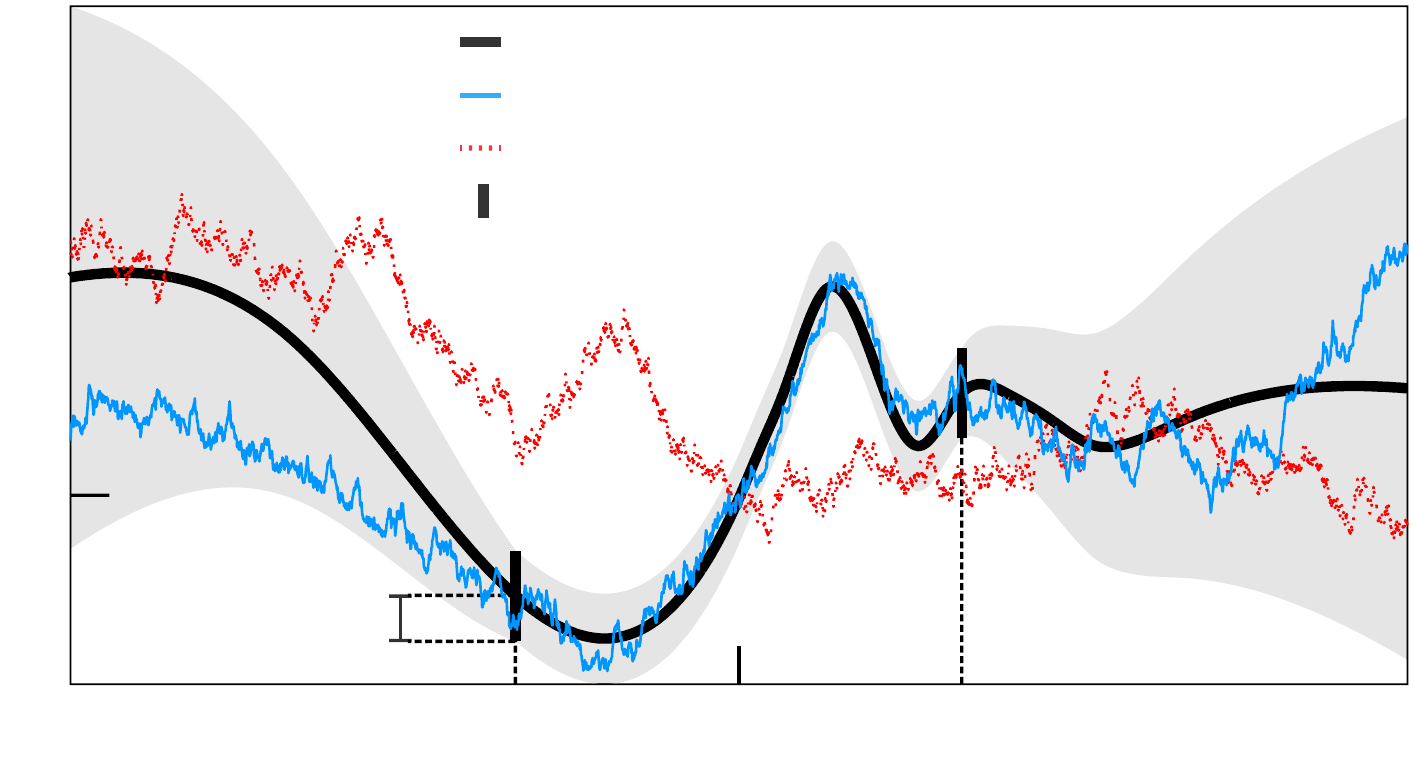} 
\put (37,50.25) {test trajectory $f$}
\put (37,46.47) {sample path in neighborhood \eqref{jointapprox}}
\put (37,42.7) {sample path out of neighborhood}
\put (37,38.7) {bottlenecks at $t=\pm T$}
\put (25.3,9.1) {$\delta$}
\put (0.2,25.2) {\rotatebox{90}{space}} \put (2.35,17.8) {$0$} 
\put (48.7,0.1) {time} \put (51.4,2.5) {$0$} \put (66.7,2.5) {$T$}  \put (33.5,2.5) {$-T$}
\end{overpic}
\caption{\label{fig1} 
\emph{Extended forgery of continuous functions.} This example depicts a test trajectory (smooth curve), its allowed neighborhood (shaded area) and two sample paths, one (solid random walk) illustrating the generic event \eqref{jointapprox} and the other (dotted random walk), the fact that arbitrary paths have low chances to enter the expanding neighborhoods through the bottlenecks.}
\end{figure}
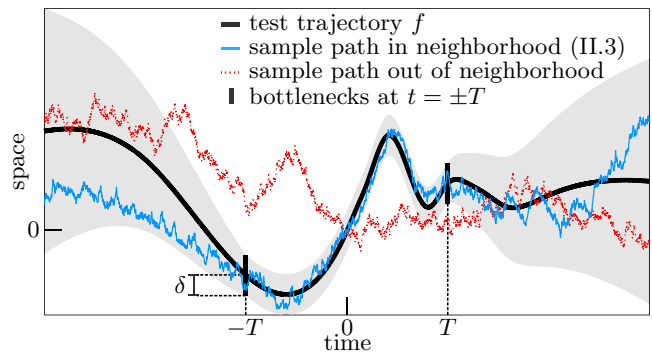
%

We now combine Levy's forgery and the asymptotic version to obtain an extension valid at all timescales.
Specifically, let us examine the \emph{joint} approximation event 
\be\label{jointapprox} \mathscr J_{f,\delta,T}=\mathscr U_{f,\delta,T}\, \textstyle\bigcap\, \mathscr E_{f,v,T}\quad \textrm{with $v=\delta/T$}\, .\ee
In words, the particle is constrained to follow closely the test trajectory for some time but is allowed afterwards to deviate slowly from it (Fig.~\ref{fig1}). 

\medskip\noindent{\textbfsf{Extended forgery theorem.}} {\it Let $f$ be bounded and continuous with $f(0)=0$, and $\delta,T>0$. Then $\prob(\mathscr J_{f,\delta,T})>0$.}

\medskip\noindent 
This result relies on the suitable integration of a ``local'' version of the theorem (see Section \ref{extendedforgeryproof}), but it can also be understood rather intuitively as follows.
Imagine for a moment that the events \eqref{unifapprox} and \eqref{linearapprox} were independent.
Their joint probability would merely be equal to the product of their marginal probabilities, which are positive by Levy's forgery and the asymptotic forgery, and genericity would then follow.
Actually they do interact because the associated neighborhoods are connected through the narrow bottlenecks at $t=\pm T$ (Fig.~\ref{fig1}) but this should only increase their joint probability, i.e., 
\be\label{jointprobbound} \prob(\mathscr J_{f,\delta,T}) > \prob(\mathscr U_{f,\delta,T})\, \prob(\mathscr E_{f,v,T})\, .\ee
The reason lies in the temporal continuity of Brownian motion.
A particle staying in the uniform neighborhood while $|t|\leq T$ necessarily passes through the bottlenecks, and is thus more likely to remain within the expanding neighborhood than arbitrary particles, which have low chances to even meet the bottlenecks (Fig.~\ref{fig1}). In other words, the proportion $\prob(\mathscr J_{f,\delta,T})/\prob(\mathscr U_{f,\delta,T})$ of sample paths $\BM\in \mathscr E_{f,v,T}$ among all those sample paths $\BM\in \mathscr U_{f,\delta,T}$ should be larger than the unconstrained probability $\prob(\mathscr E_{f,v,T})$, hence the bound \eqref{jointprobbound}.

%
\subsection{Forgery of statistical dependences}\label{statforgery}
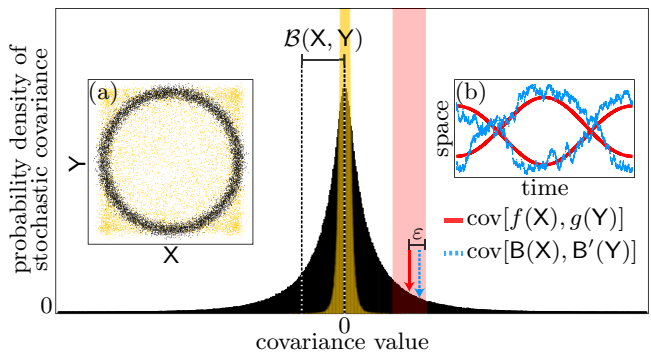
\begin{figure}
\begin{overpic}[width=8.5cm]{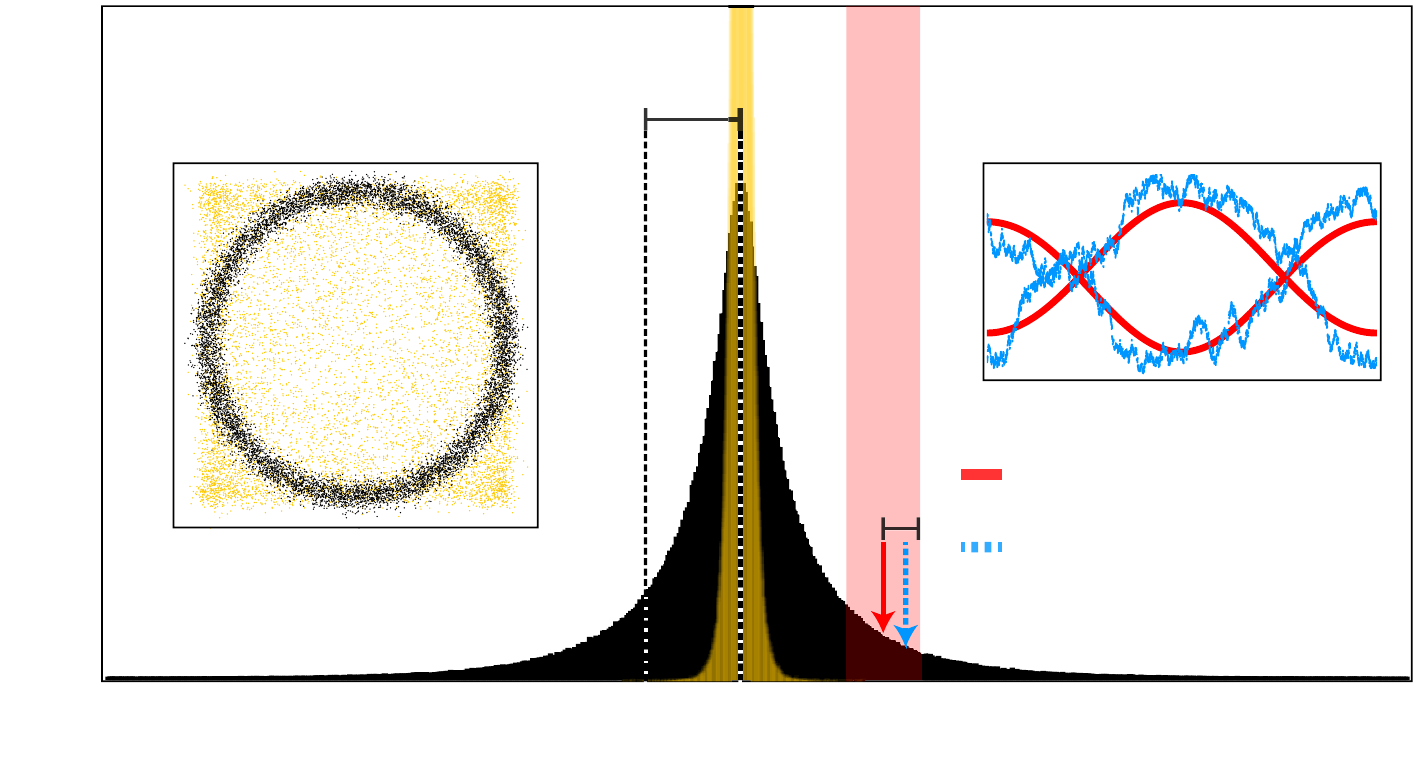}
\put (0,12) {\rotatebox{90}{probability density of}} \put (3.3,12.5) {\rotatebox{90}{stochastic covariance}} \put (4.8,5.9) {$0$} 
\put (51.45,2.8) {$0$} \put (38.5,0.3) {covariance value}
\put (42.9,47.7) {$\bcov(\X,\Y)$} 
\put (12.2,39.4) {(a)} \put (24,13.8) {$\X$} \put (9.3,28) {\rotatebox{90}{$\Y$}}
\put (69.6,39.4) {(b)} \put (67,30.4) {\rotatebox{90}{space}} \put (80,24.4) {time}
\put (63,17.5) {$\varepsilon$} 
\put (71.5,19.8) {$\cov[f(\X),g(\Y)]$} \put (71.5,14.6) {$\cov[\BM(\X),\BM'(\Y)]$} 
\end{overpic}
\caption{\label{fig2} 
\emph{Forgery of statistical dependences}. The distribution of the stochastic covariance (black histogram) and its standard deviation [$\bcov(\X,\Y)$] are shown for simulated dependent random variables [$n=10^4$ black dots, Insert (a)] as well as a test covariance (plain arrow) and a sample (dotted arrow) falling within the allowed error (shaded area). Insert (b) shows the associated functions. The case of independent variables is superimposed for comparative purposes.}
\end{figure}

%
We now turn to the analysis of statistical relations using Brownian motion. 
Let us fix two random variables $\X,\Y$ and a pair of bounded test trajectories $f,g$.
Consider the \emph{covariance} approximation event  
\be\label{covapprox}\begin{split} \mathscr C_{\X,\Y,f,g,\varepsilon}= \big\{  \big|&\cov [\BM(\X),\BM'(\Y)] \\  & - \cov[f(\X),g(\Y)]\big|<\varepsilon  \big\} \end{split}\ee
%
%
%
whereby the stochastic covariance $\cov[\BM(\X),\BM'(\Y)]$, built by picking independently two sample paths $\BM,\BM'$, coincides with the test covariance $\cov[f(\X),g(\Y)]$ up to a small error $\varepsilon>0$ (Fig.~\ref{fig2}).
We argue that this event is generic too.

The first step is to ensure that the set \eqref{covapprox} is measurable so that its probability is meaningful. 
Physically, this technical issue is rooted once again in the escape of Brownian particles to infinity. 
The stochastic covariance can be expressed as a difference of two averages $\langle \BM(\X)\,\BM'(\Y)\rangle$ and $\langle \BM(\X)\rangle\langle\BM'(\Y)\rangle$ (computed at \emph{fixed} sample paths) involving the coordinates $\BM(t),\BM'(t')$ at random moments $t=\X, t'=\Y$.
If long times and thus large coordinates are sampled too often, the two terms may diverge and lead to an ill-defined covariance, i.e., $\infty-\infty$.
To avoid this situation, we should therefore assume that asymptotic values of $\X$ and $\Y$ are unlikely enough.
Actually we shall adopt hereafter the sufficient condition that $\X,\Y$ are $L_2$, i.e., they have finite mean and variance (see \ref{measurabilityproof} for a derivation of measurability).

\medskip{\noindent\textbfsf{Forgery theorem of statistical dependences.}} {\it Let $\X,\Y$ be $L_2$ random variables, $f,g$ be boun\-ded and continuous, and $\varepsilon>0$. Then $\prob(\mathscr C_{\X,\Y,f,g,\varepsilon})>0$.}

\medskip\noindent The idea is that one way of realizing the event \eqref{covapprox} is to pick sample paths $\BM\in\mathscr J_{f,\delta,T}$, $\BM'\in\mathscr J_{g,\delta,T}$ that fit the test trajectories $f,g$ tightly ($\delta\ll 1$) over a long time period ($T\gg 1$), see Fig.~\ref{fig2}(b). 
Indeed, as shown below, we have then
\be\label{coverror} \cov[\BM(\X),\BM'(\Y)] - \cov[f(\X),g(\Y)] = \mathcal O(\delta)+\mathcal O(v) \ee
with $v=\delta/T$. 
This rough estimate explains why the event \eqref{covapprox} must occur whenever $\delta$ and $v$ are small enough [see Section \ref{statforgeryproof}, in particular Eqs.~\eqref{coverrorbound} and \eqref{bound} for a more precise error bound and the nested forgery lemma \eqref{nestedforgery} for a full derivation].
In turn, the extended forgery ensures the genericity of this selection of sample paths $\BM,\BM'$ and thus of the event $\mathscr C_{\X,\Y,f,g,\varepsilon}$ as well [the necessary condition $f(0)=g(0)=0$ can indeed be assumed without loss of generality, see Eq.~\eqref{proofcovprob2} in Section \ref{statforgeryproof}].

To understand Eq.~\eqref{coverror}, imagine first that the random times were bounded with $|\X|,|\Y|\leq T$. 
Then $\BM(\X),\BM'(\Y)$ differ from $f(\X),g(\Y)$ by less than $\delta$ for all times $\X,\Y$ [Eq.~\eqref{unifapprox}] so the covariance error must be $\mathcal O(\delta)$ at most. 
Now for unbounded random times the distance between sample paths and test trajectories may exceed $\delta$ and must actually diverge at long times, which could have led to an infinitely large covariance error if not for the fact that the occurence of $|\X|\geq T$ or $|\Y|\geq T$ is very unlikely. 
So the fit divergence $v|t|$ [see Eq.~\eqref{linearapprox}] is counterbalanced within averages by the fast decay of long times probability [e.g., $\prob(|\X|\geq |t|)\leq\langle\X^2\rangle / t^2$].
The contribution of $|\X|\geq T$ or $|\Y|\geq T$ to the covariance error is thus finite and scales as $\mathcal O(v)$, which leads to Eq.~\eqref{coverror}.

%

This forgery theorem allows us to probe enough possible relationships to establish statistical dependence or independence, as we explain now.
Consider the two probability densities of stochastic covariance shown in Fig.~\ref{fig2}, which were generated using simulations differing only by the presence or absence of coupling between $\X$ and $\Y$.
The distribution appears significantly wider for the dependent variables, so this suggests that width is the key indicator of a relation. 
Actually, for the independent variables the narrow peak observed reflects an underlying Dirac delta function (its nonzero width in Fig.~\ref{fig2} is due to finite sampling errors in the covariance estimates).
Indeed the vanishing of all stochastic covariances is a necessary condition of independence.
The impossibility of sampling nonzero values also turns out to be sufficient. 

\medskip\noindent\textbfsf{Brownian independence test.} {\it Two $L_2$ random variables $\X,\Y$ are independent iff $\cov[\BM(\X),\BM'(\Y)]=0$ {\rm a.s.}}

\medskip\noindent 
To prove sufficiency, we show that the hypothesis $\cov[\BM(\X),\BM'(\Y)]=0$ a.s.~implies that all test covariances vanish, which is equivalent to the independence of $\X$ and $\Y$ (Theorem 10.1 of Ref.~\cite{JacodProtter2003book}).
This can be understood concretely using the following thought experiment.
Imagine that $\cov[f(\X),g(\Y)]\neq 0$ for some pair of test trajectories $f,g$ and let us fix, say, $\varepsilon=|\cov[f(\X),g(\Y)]|/4$ (as in Fig.~\ref{fig2}). 
We then generate sequentially samples of stochastic covariance until the approximation event $\mathscr C_{\X,\Y,f,g,\varepsilon}$ [Eq.~\eqref{covapprox}, shaded area in Fig.~\ref{fig2}] occurs. 
The forgery of statistical dependences ensures that this sequence stops eventually and our choice of $\varepsilon$, that the last covariance sample is nonzero.
However this contradicts our hypothesis, which imposes that all trials result in vanishing covariance. (See also Section \ref{testdichotomyproof} for a set-theoretic argument.)

Figure \ref{fig2} suggests a straightforward manner to implement the Brownian independence test in practice. The sample distribution of the stochastic covariance $\cov[\BM(\X),\BM'(\Y)]$ can be generated by drawing a large number of sample paths $\BM,\BM'$, which can be approximated numerically via random walks (black histogram in Fig.~2). 
Likewise, a null distribution can be built under the hypothesis of independence, e.g., by accompanying each walk simulation with a random permutation of the sample orderings within $\X$ and $\Y$ so as to break any dependency, hence leading to a finite-sample version of the Dirac delta (yellow histogram in Fig.~2). 
These two distributions can then be compared statistically using, e.g., a Kolmogorov-Smirnov test. 
Actually, it is sufficient to focus on a comparison of their variances since width is the key parameter here.
As it turns out, this provides a more efficient implementation because it is possible to integrate out $\BM,\BM'$ analytically in the variance statistic (i.e., all sample paths are probed exhaustively without the need of random walk simulations).
This idea leads back to the notion of Brownian covariance.

\newpage

%
\subsection{Brownian covariance revisited}
%
The forgery of statistical de\-pen\-den\-ces provides an alternative approach to the theory of Brownian covariance, hereafter denoted $\bcov(\X,\Y)$. 
This dependence index emerges naturally in our context as the root mean square (r.m.s.)~of the stochastic covariance (or equivalently its standard deviation since the mean $\langle \cov[\BM(\X),\BM'(\Y)]\rangle$ vanishes identically by symmetry $\BM\longleftrightarrow -\BM$, see also Fig.~\ref{fig2}).
Thus
\be\label{bcov}  \bcov(\X,\Y)^2 = \big\langle \cov[\BM(\X),\BM'(\Y)]^2 \big\rangle\, , \ee
which is only a slight reformulation of the definition in Ref.~\cite{Szekely2009AnnAppStat}. For $L_2$ random variables $\X,\Y$, the quadratic Gaussian functional integrals over the sample paths $\BM,\BM'$ can be computed analytically and the result reduces to distance covariance, so $\bcov(\X,\Y)$ inherits all its properties \cite{Szekely2009AnnAppStat}.
Alternatively, we argue here that the central results of the theory follow in a natural manner from Eq.~\eqref{bcov}. 

The first key property is that {\it $\X,\Y$ are independent iff $\bcov(\X,\Y)=0$}.
This mirrors directly the Brownian independence test since the r.m.s.~\eqref{bcov} measures precisely the deviations of stochastic covariance from zero.
This argument thus replaces the formal manipulations on the regularized singular integrals that underlie the theory of distance covariance \cite{Szekely2007AnnStat}.
Furthermore the forgery of statistical dependences clarifies how this works physically: Brownian motion fluctuates enough to make the functional integral \eqref{bcov} probe all the possible test covariances between $\X$ and $\Y$.

The second key aspect is the straightforward sample estimation of $\mathcal B(\X,\Y)$ using a parameter-free, algebraic formula.
This is an important practical advantage over other dependency measures such as, e.g., mutual information \cite{KraskovPRE2004}. 
Instead of relying on the sample formula of distance covariance \cite{Szekely2007AnnStat, Szekely2009AnnAppStat}, Eq.~\eqref{bcov} prompts us to estimate the stochastic covariance (or rather, its square) before averaging over the Brownian paths.
So, given $n$ joint samples $\X_i,\Y_i$ ($i=1,\ldots,n$) of $\X,\Y$ and an expression for their sample covariance $\widehat{\cov}_n$, the functional integral 
\be\label{bcovsamp} \widehat{\mathcal B}_n(\X,\Y)^2 = \big\langle \widehat{\cov}_n[\BM(\X),\BM'(\Y)]^2 \big\rangle_\textrm{$(\X_i,\Y_i)_{1\leq i\leq n}$ fixed}
\ee
determines an estimator $\widehat{\mathcal B}_n(\X,\Y)$ of Brownian covariance.
If $\X,\Y$ are $L_2$ random variables, this procedure allows to build the estimation theory of Brownian (and thus, distance) covariance from that of the elementary covariance.
For instance, the rather intricate algebraic formula for the unbiased sample distance covariance \cite{Szekely2013JMultAnal, Rizzo2014AnnStat} is recovered by using, quite naturally from Eqs.~\eqref{bcov} and \eqref{bcovsamp}, an unbiased estimator $\widehat{\cov}_n[\cdot,\cdot]^2$ of the covariance squared. (See Section \ref{unbiasedBcovproof} for the explicit expressions of these  estimators and a derivation of this statement.)
The unbiasedness property $\langle \widehat{\cov}_n[\cdot,\cdot]^2\rangle=\cov[\cdot,\cdot]^2$ is then automatically transferred to the corresponding estimator \eqref{bcovsamp}, i.e., 
\be \label{unbiasednesstheorem} \big\langle\widehat{\mathcal B}_n(\X,\Y)^2\big\rangle=\mathcal B(\X,\Y)^2\, , \ee
because the $L_2$ convergence hypothesis is sufficient to ensure that averaging over the samples $\X_i,\Y_i$ commutes with the functional integration over the Brownian paths $\BM,\BM'$.

Coming back to the implementation of the Brownian independence test, once an estimator $\widehat{\mathcal B}_n(\X,\Y)$ is found, its null distribution under the hypothesis of independence should be derived for formal statistical assessment. This may be done using large $n$ approximations to obtain parametrically the asymptotic distribution or nonparametric methods also valid at small $n$ (such as sample ordering permutations, as suggested above). Asymptotic tests are derived explicitly in Refs.~\cite{Szekely2007AnnStat,Szekely2009AnnAppStat,Szekely2013JMultAnal}, where both parametric and nonparametric approaches are also illustrated on examples motivating the usefulness of this nonlinear correlation index in data analysis (including comparisons to linear correlation tests and assessments of statistical power).

It is noteworthy that our construction of Brownian covariance and its estimator can be generalized by formally replacing the Brownian paths $\BM,\BM'$ with other stochastic processes or fields (in which case we may consider multivariate variables $\X,\Y$).
This determines a simple rule to engineer a wide array of dependence measures via functional integrals, and opens the question of what processes allow to characterize independence.
Our approach relied on the ability to probe generically all possible test covariances but, critically, the class of processes satisfying a forgery of statistical dependences might be relatively restricted.
On the other hand, the original theory of Brownian covariance does extend to multidimensional Brownian fields or fractional Brownian motion \cite{Szekely2009AnnAppStat}, which are not continuous or Markovian (two properties central for the forgery theorems).
So the forgery of statistical dependences provides a new and elegant tool to establish independence, but it may only represent a particular case of a more general theory of functional integral-based correlation measures.

%
\section{MATHEMATICAL ANALYSIS}\label{mathematicalconsiderations}
\setcounter{equation}{0}

We now proceed with a more detailed examination of our results. 
The most technical parts of the proofs are relegated to \ref{technicalappendix}.

%
\subsection{Asymptotic forgery and duality}\label{dualityproof}
We sketched in Section \ref{extendedforgery} a derivation of the asymptotic forgery theorem using the law of large numbers.
A generalization can actually be developed fully (see \ref{strongforgery}).
However, this particular case enjoys a concise proof based on a symmetry argument.

\medskip\noindent\textbfsf{The short/long time duality.} We start by establishing the duality relation
\be\label{duality} \prob(\mathscr E_{f,v,T}) = \prob(\mathscr U_{f_\textsc{d},v,1/T})\, , \ee
where the dual $f_\textsc{d}$ of the function $f$ is given by
\be\label{fdual} f_\textsc{d}(t) = \begin{cases*} |t| f(1/t) & for $t\neq0\, ,$ \\ 0 & for $t=0\, .$ \end{cases*}\ee
%

\noindent\textbfsf{Proof.}
By the time-inversion symmetry of Brownian motion \cite{Freedman1983book}, replacing $\BM(t)$ by $|t|\BM(1/t)$ in the right-hand side of Eq.~\eqref{linearapprox} determines a new event with the same probability as $\mathscr E_{f,v,T}$. 
Explicitly, this event is
\be\label{dualevent}\begin{split} \big\{  &\big||t|\BM(1/t)-f(t)\big| <v |t| \, , \,  \forall\, |t|\geq T \big\} \\ &= \big\{  |\BM(t')-f_\textsc{d}(t')| <v \, , \,  \forall\, 0<|t'|\leq 1/T \big\}\, , \end{split}\ee
where $t'=1/t$ and Eq.~\eqref{fdual} were used. 
The condition at $t'=0$ holds identically since $\BM(0)=f_{\textsc{d}}(0)=0$, so \eqref{dualevent} coincides with $\mathscr U_{f_\textsc{d},v,1/T}$.
This yields Eq.~\eqref{duality}. 
\qed

\medskip\noindent\textbfsf{Proof of the asymptotic forgery theorem.} 
The theorem naturally follows from this duality and Levy's forgery.
The explicit formula \eqref{fdual} establishes that $f_\textsc{d}$ is continuous [for $t\neq 0$ this corresponds to the continuity of $f$, and for $t=0$ to its boundedness $\sup_{t\in\mathbb R}|f(t)|\leq M$ since then $|t f(1/t)|\leq M|t|\rightarrow 0$ as $t\rightarrow 0$]. 
Levy's forgery theorem thus applies and shows that the right-hand side of Eq.~\eqref{duality} is nonzero. 
\qed

\subsection{Local extended forgery}\label{extendedforgeryproof}

We now describe an analytical derivation of the extended forgery theorem that formalizes the intuitive argument given in Section \ref{extendedforgery}.
The ensuing bound for the probability of the joint event \eqref{jointapprox} does not quite reach that in Eq.~\eqref{jointprobbound} but is sufficient to ensure genericity.

\medskip\noindent\textbfsf{Restriction to positive-time events.}
As a preliminary, it will be useful to consider the positive- and negative-time events $\mathscr U_{f,\delta,T}^\pm$ and $\mathscr E_{f,v,T}^\pm$, which correspond to \eqref{unifapprox} and \eqref{linearapprox} except that their defining conditions are enforced only for $0\leq \pm t\leq T$ and $\pm t\geq T$, respectively.
These events are generic too because they are less constrained (e.g., $\mathscr U_{f,\delta,T}^\pm$ contains $\mathscr U_{f,\delta,T}=\mathscr U_{f,\delta,T}^+\bigcap\mathscr U_{f,\delta,T}^-$) so that
\begin{align}\label{Levy+} \prob(\mathscr U_{f,\delta,T}^\pm) &\geq \prob(\mathscr U_{f,\delta,T}) >0\, , \\ \label{asymptotic+} \prob(\mathscr E_{f,v,T}^\pm) &\geq \prob(\mathscr E_{f,v,T}) >0\, ,\end{align}
by monotonicity of the probability measure $\prob(\cdot)$, Levy's forgery, and the asymptotic forgery.

We are going to focus below on the derivation of  
\be\label{fwdtimeextendedforgery} \prob(\mathscr U_{f,\delta,T}^+\, \textstyle\bigcap\, \mathscr E_{f,v,T}^+)>0\, . \ee
This is sufficient to prove the extended forgery because the backward-time ($t\leq 0$) part of Brownian motion is merely an independent copy of its forward-time ($t\geq 0$) part.
Hence a similar result necessarily holds for the negative-time events and, in turn,
\be\label{fullextendedforgery} \prob(\mathscr J_{f,\delta,T}) = {\textstyle\prod_{\sigma=\pm}}\, \prob(\mathscr U_{f,\delta,T}^\sigma\, \textstyle\bigcap\, \mathscr E_{f,v,T}^\sigma)>0\, .\ee
%

\noindent\textbfsf{The integral formula.} The first step in the demonstration of Eq.~\eqref{fwdtimeextendedforgery} relies on the following explicit expression for the joint probability as a functional integral.
Let us introduce the family of \emph{auxiliary} events
\be\label{defAx}\mathscr A_{f,\delta,T}^x = \big\{ |\BM(t)+x-f(T+t)|<\delta+vt \, , \,  \forall\, t\geq 0  \big\} \ee
and denote their probability by
\be \label{psidef} p_{f,\delta,T}(x) = \prob(\mathscr A_{f,\delta,T}^x) \ee 
for all $x\in\mathbb R$.
Then 
\be\label{integral} \prob(\mathscr U_{f,\delta,T}^+ \ {\textstyle\bigcap} \ \mathscr E_{f,v,T}^+) = \big\langle \boldsymbol 1_{\mathscr U_{f,\delta,T}^+}\, p_{f,\delta,T}\big(\BM(T)\big) \big\rangle\, . \ee
The first factor in this expectation value denotes the indicator function of the event $\mathscr U_{f,\delta,T}^+$ and enforces the constraint that all considered sample paths $\BM$ must lie within the uniform neighborhood \eqref{unifapprox} for $0\leq t\leq T$. 
In the second factor the function \eqref{psidef} is evaluated at the position of the random path $\BM(t)$ at $t=T$. 
As we explain below this function is Borel measurable, so $p_{f,\delta,T}(\BM(T))$ represents a proper random variable and the expectation value is well defined.

\medskip\noindent\textbfsf{Proof.} 
The formula \eqref{integral} is a direct consequence of the two following statements, each involving the conditional probability $\prob[\mathscr E_{f,v,T}^+\, | \, \BM(T)]$ that the event $\mathscr E_{f,v,T}^+$ occurs under the constraint that the Brownian motion passes at time $t=T$ through a given location $x$ randomly distributed as $\BM(T)$:
\be\label{markovintegral} \prob(\mathscr U_{f,\delta,T}^+ \ {\textstyle\bigcap} \ \mathscr E_{f,v,T}^+) = \big\langle \boldsymbol 1_{\mathscr U_{f,\delta,T}^+}\, \prob[\mathscr E_{f,v,T}^+\, | \, \BM(T)] \big\rangle\, , \ee
and
\be \label{condtoBT} \prob[\mathscr E_{f,v,T}^+\, | \, \BM(T)] = p_{f,\delta,T}\big(\BM(T)\big) \quad \textrm{a.s.} \ee
They follow from fairly standard arguments about Brownian motion \cite{Freedman1983book,GikhmanSkorokhod1969book} that we detail in \ref{technicalappendix-markov} and \ref{technicalappendix-condtoBT}.
The first implements the Markov property that $\mathscr E_{f,v,T}^+$ depends on its past only via $\BM(t)$ at the boundary time $t=T$.
The second provides an explicit representation of the random variable $\prob[\mathscr E_{f,v,T}^+\, | \, \BM(T)]$ (see also Chapter 1, Theorem 12 in Ref.~\cite{Freedman1983book}).
\qed

The next step is to show that the integrand in the right-hand side of Eq.~\eqref{integral} cannot vanish identically, which provides a ``local'' version of the extended forgery. 
The full theorem will follow by integration.

\medskip\noindent\textbfsf{Extended forgery theorem (local version).} {\it There exists a subinterval $J$ of the bottleneck $f(T)-\delta<x<f(T)+\delta$ at $t=T$ such that
\be\label{Levyloc} \prob\big(\mathscr U_{f,\delta,T}^+\, \textstyle\bigcap\, \{ \BM(T)\in J\}\big) \geq \prob(\mathscr U_{g,\ell,T}^+)\ee
for some continuous function $g$ satisfying $g(0)=0$ and some distance parameter $\ell>0$, and
\be\label{asymptoticloc} p_{f,\delta,T}(x) > \prob(\mathscr E_{f,v,T}^+)\ee
%
for all $x\in J$.}

By Eqs.~\eqref{Levy+} and \eqref{asymptotic+}, both lower bounds are positive. 
The two parts of the theorem are closely related to Levy's forgery and the asymptotic forgery, and we treat them separately. 

\medskip\noindent\textbfsf{Proof of \eqref{Levyloc}.}
This property actually holds for arbitrary subintervals $J$, which we shall choose open and parameterized as $x_0-\ell < x < x_0+\ell$ with $|x_0-f(T)|\leq\delta-\ell$ to ensure inclusion in the bottleneck.
We also define the continuous function $g$ by $g(t)=f(t)+t[x_0-f(T)]/T$.
This setup ensures that $\mathscr U_{g,\ell,T}^+\subset \mathscr U_{f,\delta,T}^+$ and $\mathscr U_{g,\ell,T}^+\subset \{ \BM(T)\in J \}$, as the condition $|\BM(t) - g(t)|<\ell$ for all $0\leq t\leq T$ implies
\begin{align} \nonumber |\BM(t) - f(t)| & \leq |\BM(t) - g(t)| +  |g(t) - f(t)| \\ & < \ell + |x_0-f(T)| \leq\delta\, \end{align}
and $|\BM(T)-x_0|=|\BM(T)-g(T)|<\ell$ as $g(T)=x_0$.
The bound \eqref{Levyloc} follows from these two inclusions by the monotonicity of $\prob(\cdot)$.
\qed

\medskip\noindent\textbfsf{Two properties of the function $\boldsymbol{p_{f,\delta,T}}$.} 
For the second part it is useful to interject here the following simple statements about the function \eqref{psidef}:
\begin{enumerate}[(i)]
\itemsep -0.3em
\item\label{p1} $p_{f,\delta,T}$ vanishes identically outside the bottleneck,
\item\label{p2} $p_{f,\delta,T}$ is continuous within the bottleneck (and therefore, it is Borel measurable as well).
\end{enumerate}

The first claim rests on the observation that the set \eqref{defAx} is empty whenever $|x-f(T)|\geq \delta$ (since its defining condition at $t=0$ cannot be satisfied, as $\BM(0)=0$), so we find $p_{f,\delta,T}(x)=\prob(\emptyset)=0$. 

The second claim may appear quite clear as well, since the set \eqref{defAx} should vary continuously with $x$, but this intuition is not quite right. 
For a more accurate statement, let us fix a point $x$ in the bottleneck and consider an arbitrary sequence $x_n$ converging to it. 
Then the limit event of $\mathscr A_{f,\delta,T}^{x_n}$ (assuming it exists) coincides with $\mathscr A_{f,\delta,T}^{x}$ modulo a zero-probability set, so by continuity of $\prob(\cdot)$
\be\label{Pcontinuous} \lim_{n\rightarrow\infty} p_{f,\delta,T}(x_n) = p_{f,\delta,T}(x)\, .\ee
The full proof appears rather technical, so we only sketch the key ideas here and relegate the details to \ref{technicalappendix-continuity}.
The limit event imposes that sample paths lie within the expanding neighborhood associated with \eqref{defAx} but are allowed to reach its boundary [essentially because the large $n$ limit of the strict inequalities \eqref{defAx} at $x=x_n$ yields a nonstrict inequality].
However the latter hitting event a.s.~never happens because the typical roughness of Brownian motion forbids meeting a boundary curve without crossing it and thus, leaving the neighborhood (see \ref{crossinglaw} for this ``boundary-crossing law'', which generalizes Lemma 1 on page 283 of Ref.~\cite{GikhmanSkorokhod1969book} to time-dependent levels.).

\medskip\noindent\textbfsf{Proof of \eqref{asymptoticloc}.}
The key observation here is that
\be\label{tutu2}\prob(\mathscr E_{f,v,T}^+) = \int_{-\infty}^\infty \mathrm{d} x\, \frac{\exp(-x^2/2T)}{\sqrt{2\pi T}}\,  p_{f,\delta,T}(x) \, ,\ee
%
which is merely the integral formula \eqref{integral} with the constraint $\BM\in\mathscr U_{f,\delta,T}^+$ removed [equivalently, this is Eq.~\eqref{tutu} with $S=\mathbb R$] and the expectation over $\BM(T)$ made explicit. 
If $p_{f,\delta,T}(x)\leq\prob(\mathscr E_{f,v,T}^+)$ everywhere, then for almost all $x$ this inequality must saturate to an equality by Eq.~\eqref{tutu2} and thus $p_{f,\delta,T}(x)>0$ by Eq.~\eqref{asymptotic+}. 
This contradicts property \eqref{p1} of $p_{f,\delta,T}$, so that Eq.~\eqref{asymptoticloc} must hold for at least one point $x$ in the bottleneck, and thus also on some subinterval $J$ by the continuity property \eqref{p2}.
\qed

\medskip\noindent\textbfsf{Proof of the extended forgery theorem.} 
Finally we combine the local version of the theorem with the integral formula \eqref{integral} to obtain
\be \label{lastbound} \prob\big(\mathscr U_{f,\delta,T}^+\,\textstyle\bigcap\, \mathscr E_{f,v,T}^+\big) > \prob(\mathscr U_{g,\ell,T}^+)\, \prob(\mathscr E_{f,v,T}^+)\, .\ee
Indeed, further constraining the paths to $\BM(T)\in J$ allows to bound the right-hand side of \eqref{integral} from below by 
\begin{align} \nonumber\big\langle \boldsymbol 1&_{\mathscr U_{f,\delta,T}^+\, \bigcap\,\{\BM(T)\in J\}}\, p_{f,\delta,T}\big(\BM(T)\big) \big\rangle \\ \nonumber & > \prob\big(\mathscr U_{f,\delta,T}^+\, \textstyle\bigcap\, \{ \BM(T)\in J\}\big)\, \prob(\mathscr E_{f,v,T}^+) & \textrm{by \eqref{asymptoticloc},} \\ \nonumber & \geq \prob(\mathscr U_{g,\ell,T}^+)\, \prob(\mathscr E_{f,v,T}^+) & \textrm{by \eqref{Levyloc}.} \end{align}
%
%
%
In passing through the first inequality we also used the identity $\langle \boldsymbol 1_{\mathscr M}\rangle=\prob(\mathscr M)$ valid for any event $\mathscr M$.
The theorem \eqref{fwdtimeextendedforgery} follows from the inequality \eqref{lastbound} together with Eqs.~\eqref{Levy+} and \eqref{asymptotic+}. 
\qed

It is noteworthy that we stated and proved the extended forgery under the assumption that $v=\delta/T$ for convenience, but in the above arguments this restriction was actually artificial so the theorem holds for arbitrary parameters $\delta,v,T>0$. 

\subsection{Nested forgery of statistical dependences}\label{statforgeryproof}

We showed in Section \ref{statforgery} that the forgery of statistical dependences is induced by the extended forgery using the somewhat rough estimate \eqref{coverror} of the covariance error.
We provide now an exact upper bound and a complete proof of the theorem.

\medskip\noindent\textbfsf{Error bound.}
For arbitrary sample paths $\BM\in \mathscr J_{f,\delta,T}$ and $\BM'\in \mathscr J_{g,\delta,T}$, we have
\be\label{coverrorbound} \big| \cov[\BM(\X),\BM'(\Y)] - \cov[f(\X),g(\Y)] \big| < \varepsilon_{\textsc{b} }(\delta,v)\ee
where $v=\delta/T$ and the error bound $\varepsilon_{\textsc{b}}(\delta,v)$ is given by 
%
%
%
\be\label{bound} \begin{split} \varepsilon_{\textsc b}(\delta,v) = \ &  2(M_f+M_g) \delta +2\delta^2\\ &+2\big(M_g\langle|\X|\rangle + M_f\langle|\Y|\rangle \big) v \\ &+2\langle|\X|+|\Y|\rangle \delta v \\ &+\big(\langle|\X\Y|\rangle+\langle|\X|\rangle\langle|\Y|\rangle\big) v^2\, , \end{split}\ee
with $M_f=\sup_{t\in\mathbb R} |f(t)|$ and $M_g=\sup_{t\in\mathbb R} |g(t)|$.

The derivation of this estimate relies on a relatively straightforward application of a series of inequalities and is relegated to \ref{technicalappendix-bound}.
Its significance rests on the fact that, when the polynomial coefficients in \eqref{bound} are finite, the covariance error can be made arbitrarily small by taking $\delta,v\rightarrow 0$. 
Hence we obtain the following key intermediate result.

\medskip\noindent\textbfsf{The nested forgery lemma.} {\it Assume that $M_f$, $M_g$, $\langle |\X|\rangle$, $\langle |\Y|\rangle$, and $\langle |\X\Y|\rangle$ are finite, and let $\varepsilon>0$. Then there exists $\delta,T>0$ such that
\be\label{nestedforgery} \mathscr J_{f,\delta,T} \times \mathscr J'_{g,\delta,T} \subset \mathscr C_{\X,\Y,f,g,\varepsilon}\, , \ee
where the prime on event $\mathscr J_{g,\delta,T}$ indicates that it applies to the process $\BM'$.}

\medskip\noindent\textbfsf{Proof of the forgery theorem of statistical dependences.} 
This is a direct corollary. 
The $L_2$ hypothesis and the boundedness assumption on $f,g$ ensure that the lemma \eqref{nestedforgery} applies. 
Using the monotonicity of $\prob(\cdot)$ and the independence of $\BM,\BM'$ then yields
%
%
%
\begin{align}\prob(\mathscr C_{\X,\Y,f,g,\varepsilon})&\geq \prob(\mathscr J_{f,\delta,T}\times\mathscr J'_{g,\delta,T})\nonumber \\ \label{proofcovprob1} &=\prob(\mathscr J_{f,\delta,T})\, \prob(\mathscr J_{g,\delta,T})\, .\end{align}
It is now tempting to invoke the extended forgery theorem, but here the conditions $f(0)=g(0)=0$ were not imposed.
This is however not an issue thanks to the translation invariance of covariance, i.e., $\cov[f(\X),g(\Y)]$ is equal to $\cov[f(\X)-f(0),g(\Y)-g(0)]$. 
Thus
\begin{align} \label{proofcovprob2} \prob(\mathscr C_{\X,\Y,f,g,\varepsilon})& =\prob(\mathscr C_{\X,\Y,f-f(0),g-g(0),\varepsilon})  \\ &\geq\prob(\mathscr J_{f-f(0),\delta,T})\, \prob(\mathscr J_{g-g(0),\delta,T}) \nonumber >0\, ,\end{align}
where we used Eq.~\eqref{proofcovprob1} and the extended forgery theorem  in the second line.
\qed

Note that the $L_2$ assumption used in Section \ref{statforgery} was slightly stronger than needed, as made clear by the nested forgery lemma.

\subsection{Brownian independence test and dichotomy}\label{testdichotomyproof}

To prove that the assertion $\cov[\BM(\X),\BM'(\Y)]=0$ a.s.~implies $\cov[f(\X),g(\Y)]=0$, we used in Section \ref{statforgery} a discrete sampling method for which it is straightforward that the covariance approximation event \eqref{covapprox} must occur simultaneously with the \emph{zero} covariance event
\be\label{zerocov} \mathscr Z_{\X,\Y}=\big\{ \cov[\BM(\X),\BM'(\Y)]=0 \big\}\, . \ee 
Their compatibility, which was indeed central in our derivation, is actually a general property of probability theory and we use it here to provide an alternative, set-theoretic argument.

\medskip\noindent\textbfsf{The dichotomy lemma.} {\it For arbitrary functions $f,g$ and parameter $\varepsilon$, we have}
\be\label{dichotomy} \mathscr C_{\X,\Y,f,g,\varepsilon}\textstyle\bigcap\mathscr Z_{\X,\Y}=\begin{cases*} \mathscr Z_{\X,\Y} & if $|\cov[f(\X),g(\Y)]|<\varepsilon$ \\ \emptyset & otherwise. \end{cases*}
\ee

\medskip\noindent\textbfsf{Proof.} It is a direct consequence of Eqs.~\eqref{covapprox} and \eqref{zerocov} that the intersection is characterized equivalently by the two conditions $\cov[\BM(\X),\BM'(\Y)]=0$ and $|\cov[f(\X),g(\Y)]|<\varepsilon$. 
\qed

\medskip\noindent\textbfsf{Second proof of sufficiency for the Brownian independence test.} By the forgery of statistical dependences the event $\mathscr C_{\X,\Y,f,g,\varepsilon}$ is generic for $f,g$ bounded and continuous and $\varepsilon>0$, and by hypothesis the event $\mathscr Z_{\X,\Y}$ occurs a.s.
Since the intersection of a generic event and an almost sure event is never empty [if it were empty, the generic event would be a subset of a zero probability event (i.e., the complement of the almost sure event), which is forbidden by monotonicity of $\prob(\cdot)$], the second possibility in Eq.~\eqref{dichotomy} is ruled out so $|\cov[f(\X),g(\Y)]|<\varepsilon$ must hold true. 
Since $\varepsilon>0$ was arbitrary, we conclude that $\cov[f(\X),g(\Y)]=0$.
\qed

\subsection{Unbiased estimation of Brownian covariance}\label{unbiasedBcovproof}

We now exemplify how an explicit estimator of Brownian covariance can be derived from the functional integral \eqref{bcovsamp}. 
Our construction enforces unbiasedness at finite sampling and allows to recover the unbiased sample formula of distance covariance \cite{Szekely2013JMultAnal, Rizzo2014AnnStat} that we review first.

\medskip\noindent\textbfsf{The unbiased estimator.} 
Let us introduce the distance $a_{ij}=|\X_i-\X_j|$ between the samples $\X_i$ of $\X$ as well as its ``U-centered'' version
\be\label{ucentering} A_{ij}=a_{ij}-\frac{\sum_{k} (a_{ik}+a_{kj})}{n-2}+\frac{\sum_{k,l} a_{kl}}{(n-1)(n-2)}\, \cvp \ee
where $1\leq i,j,k,l\leq n$.
The analogous matrices for the corresponding samples of $\Y$ are denoted $b_{ij}$ and $B_{ij}$, respectively.
With these notations and assuming $n\geq 4$,
\be\label{bcovsampunbiased}  \widehat{\bcov}_n(\X,\Y)^2=\frac{1}{4n(n-3)} \sum_{\substack{\textrm{distinct} \\ i,j}} A_{ij} B_{ij}\, . \ee
This expression differs from the formula given in Refs.~\cite{Szekely2013JMultAnal, Rizzo2014AnnStat} by a trivial factor $1/4$ due to our use of the standard normalization for Brownian motion.
The asymptotic distribution of this estimator under the null hypothesis of independence was worked out in Ref.~\cite{Szekely2013JMultAnal}.
The unbiasedness property \eqref{unbiasednesstheorem} of Eq.~\eqref{bcovsampunbiased} was also proven in Refs.~\cite{Szekely2013JMultAnal,Rizzo2014AnnStat}, but here it will follow directly from our derivation based on the functional integral \eqref{bcovsamp}, which starts naturally from an unbiased estimation of the covariance squared.

\medskip\noindent\textbfsf{Estimation of covariance squared.} 
It is convenient to introduce the new random variables $\x=\BM(\X),\y=\BM'(\Y)$ and their samples $\x_i=\BM(\X_i),\y_i=\BM'(\Y_i)$, all being defined at fixed sample paths $\BM,\BM'$.
An unbiased estimator for $\cov(\x,\y)$ itself is well-known from elementary statistics, but it can be checked by developing its square that the estimation of $\cov(\x,\y)^2$ is then hampered by systematic errors of order $1/n$. 
Here, given $n\geq 4$, we shall rather define
\be\label{estimcov2}\widehat{\cov}_n(\x,\y)^2= \frac{(n-4)!}{n!} \sideset{}{'}\sum \x_i\x_j \big(\y_i \y_j -\y_i\y_k-\y_j\y_k+\y_k\y_l \big)\, , \ee
where the primed sum is taken over all distinct indices $1\leq i,j,k,l\leq n$. 
This expression differs from the aforementioned development by $\mathcal O(1/n)$ and is indeed free from finite sampling biases.

This can be proven by averaging Eq.~\eqref{estimcov2} over the $n$ joint samples $\x_i,\y_i$. 
Indeed, using the identities $\langle \x_i \x_j \y_i \y_j \rangle=\langle \x\y\rangle^2$, $\langle\x_i \x_j\y_i\y_k\rangle=\langle \x\y\rangle\langle\x\rangle\langle\y\rangle$, and $\langle\x_i \x_j\y_k\y_l\rangle=\langle\x\rangle^2\langle\y\rangle^2$ and the fact that the primed sum contains $n!/(n-4)!$ terms, we find
%
%
%
\begin{align} \nonumber\big\langle\widehat{\cov}_n(\x,\y)^2\big\rangle & =  \langle \x\y\rangle^2-2\langle \x\y\rangle\langle \x\rangle\langle\y\rangle+\langle \x\rangle^2\langle\y\rangle^2 \\ \label{unbiasednessofcov2}& = \cov(\x,\y)^2\, . \end{align}

\medskip\noindent\textbfsf{Derivation of Eq.~\eqref{bcovsampunbiased} from Eq.~\eqref{bcovsamp}.} 
We now average Eq.~\eqref{estimcov2} with $\x_i=\BM(\X_i)$ and $\y_i=\BM'(\Y_i)$ over the independent random paths $\BM,\BM'$, while the samples $\X_i,\Y_i$ are being kept constant.
The computation factors into the functional integration of $\x_i\x_j=\BM(\X_i)\BM(\X_j)$ and $\y_i\y_j=\BM'(\Y_i)\BM'(\Y_j)$, so $\widehat\bcov_n(\X,\Y)^2$ is obtained from the right-hand side of Eq.~\eqref{estimcov2} by replacing $\x_i\x_j$ and $\y_i\y_j$ with the autocorrelation functions \cite{Freedman1983book}
\be\label{BMautocorr} \begin{split} \big\langle \BM(\X_i) \BM(\X_j) \big\rangle &=\tfrac{1}{2} \big(|\X_i|+|\X_j|-a_{ij}\big)\, , \\ \big\langle \BM'(\Y_i) \BM'(\Y_j) \big\rangle &=\tfrac{1}{2} \big(|\Y_i|+|\Y_j|-b_{ij}\big)\, , \end{split} \ee
respectively.
This substitution rule can actually be simplified further to $\x_i\x_j\rightarrow -a_{ij}/2$ because the terms involving $|\X_i|$ cancel out thanks to the algebraic identity
\be\label{cancelout} \sum_{\substack{\textrm{all distinct} \\ j,k,l\neq i}} (\y_i \y_j -\y_i\y_k-\y_j\y_k+\y_k\y_l)=0\, . \ee
Similar cancellations also allow to use $\y_i\y_j\rightarrow -b_{ij}/2$.
We thus obtain the unbiased estimator
\be\label{bcovsampunbiased0} \widehat{\bcov}_n(\X,\Y)^2= \frac{(n-4)!}{4n!} \sideset{}{'}\sum a_{ij}\, ( b_{ij}  - b_{ik} - b_{kj} + b_{kl} )\, .\ee

The equivalence with Eq.~\eqref{bcovsampunbiased} is not obvious at first sight. 
To make contact with it, let us fix $i,j$ and consider the sum of the second factor over $k,l$, all indices being distinct. 
This contribution is the sum of the following four terms:
\begin{align*} \textstyle{\sum_{k,l}'} b_{ij} & = (n-2)(n-3) b_{ij}\, , \\  -\textstyle{\sum_{k,l}'} b_{ik} & 
= (n-3) \big(\textstyle{b_{ij}-\sum_{k}} b_{ik}\big)\, , \\ -\textstyle{\sum_{k,l}'} b_{kj} & 
= (n-3) \big(b_{ij} - \textstyle{\sum_{k}} b_{kj}\big)\, , \\ \textstyle{\sum_{k,l}'} b_{kl} & = \textstyle{\sum_{k,l}} b_{kl} - 2\textstyle{\sum_{k}} (b_{ik}+b_{kj}) +2 b_{ij}\, ,
\end{align*}
where the sums in the right-hand sides now run over unconstrained indices $1\leq k,l \leq n$.
The total thus reduces to $(n-1)(n-2)B_{ij}$, so Eq.~\eqref{bcovsampunbiased0} can be rewritten as 
\be\label{bcovsampunbiased1}  \widehat{\bcov}_n(\X,\Y)^2=\frac{1}{4n(n-3)} \sum_{\substack{\textrm{distinct} \\ i,j}} a_{ij} B_{ij}\, . \ee
Finally we recover Eq.~\eqref{bcovsampunbiased} because $a_{ij}$ can be replaced by its U-centering $A_{ij}$. 
Indeed the extra terms cancel in the sum thanks to the centering property $\sum_{j\neq i} B_{ij}=0$ (for all $i$ fixed), which can be checked from the definition \eqref{ucentering}.


\section*{Acknowledgments}
This research was carried over in the context of methodological developments for the MEG project at CUB -- H\^opital Erasme (Universit\'e libre de Bruxelles, Brussels, Belgium), which is financially supported by the Fonds Erasme (convention ``Les Voies du Savoir'', Fonds Erasme, Brussels, Belgium).

%
\appendix
\renewcommand\thesection{Appendix \Alph{section}}

%
\section{A STRONGER VERSION OF THE FORGERY THEOREMS} \label{strongforgery}
\renewcommand\theequation{\Alph{section}.\arabic{equation}}
\setcounter{equation}{0}

%

The various forgery theorems obtained by extending Levy's original forgery relied on approximations involving neighborhoods expanding at long times with a constant speed $v$. 
However it turns out that this assumption can be weakened. 
We now describe a stronger version of the asymptotic forgery theorem and then briefly discuss some implications.

\medskip\noindent\textbfsf{Slowly expanding neighborhoods.} 
The key observation is that, in our discussion of the asymptotic forgery, the law of large numbers can be replaced almost verbatim by the more precise law of the iterated logarithm, which establishes that Brownian motion diverges a.s.~at long times not faster than $\sqrt{2|t| \log \log |t|}$ \cite{Freedman1983book}.

Thus we naturally generalize the definition \eqref{linearapprox} using neighborhoods that expand strictly faster than almost all Brownian paths. 
Explicitly, we let
\be\label{nonlinearapprox} \mathscr E_{f,\phi,T} = \big\{  |\BM(t)-f(t)| <\phi(|t|) \, , \,  \forall\, |t|\geq T  \big\}\, , \ee
where the growth function $\phi(t)$ is positive and continuous on $t>0$, and satisfies the divergence condition
\be\label{growthfunc}  \lim_{t\rightarrow+\infty} \frac{\sqrt{t \log\log t}}{\phi(t)} =0\, . \ee
The particular example $\phi(t)=v t$ leads back to Eq.~\eqref{linearapprox}.

The asymptotic behavior \eqref{growthfunc}, together with the law of the iterated logarithm, indeed ensures that the expansion described by $\phi(t)$ is a.s.~faster than Brownian motion, i.e.,
\be\label{Bphitozero}  \lim_{|t|\rightarrow\infty} \frac{\BM(t)}{\phi(|t|)} =0\quad \textrm{a.s.}\, , \ee
a fact that we shall use momentarily. 
To prove this assertion, let us start from the inequality
\begin{equation} \begin{split} \limsup_{|t|\rightarrow\infty} \frac{|\BM(t)|}{\phi(|t|)} \leq \ & \limsup_{|t|\rightarrow\infty} \frac{|\BM(t)|}{\sqrt{2|t|\log\log|t|}} \\ & \times \limsup_{|t|\rightarrow\infty} \frac{\sqrt{2|t|\log\log|t|}}{\phi(|t|)}\, \cvd\end{split}\end{equation}
The law of the iterated logarithm states precisely that the first factor in the right-hand side equals a.s.~to one, and the second factor tends to zero by definition \eqref{growthfunc}.
Therefore we obtain $\limsup_{|t|\rightarrow\infty} |\BM(t)|/\phi(|t|)=0$ a.s., which is equivalent to Eq.~\eqref{Bphitozero}.

\medskip\noindent{\textbfsf{Strong asymptotic forgery theorem.}} {\it Let $f$ be bounded and continuous, $\phi$ be a growth function as specified above, and $T>0$. Then $\prob(\mathscr E_{f,\phi,T})>0$.}

\medskip\noindent\textbfsf{Proof.} 
We revisit here the strategy of proof sketched in the main text. 
Let us start by demonstrating that 
\be \label{largeTlim} \lim_{T\rightarrow\infty} \prob(\mathscr E_{f,\phi,T}) = 1\, .\ee
The event \eqref{nonlinearapprox} increases monotonously when $T$ is raised (since the defining condition becomes less strict) so $\lim_{T\rightarrow\infty} \prob(\mathscr E_{f,\phi,T})$ is equal to $\prob(\bigcup_{T>0}\,\mathscr E_{f,\phi,T})$ by continuity of the measure $\prob(\cdot)$. 
Therefore we need to show that the limit event $\bigcup_{T>0}\,\mathscr E_{f,\phi,T}$ occurs a.s.
This is the set-theoretic version of the statement that, for almost all sample paths $\BM$, we can find some $T>0$ ensuring the inequality $|\BM(t)-f(t)|/\phi(|t|)<1$, $\forall\, |t|\geq T$. 
In turn this is a particular case of the slightly stronger assertion
\be\label{iteratedlogs} \lim_{|t|\rightarrow\infty} \frac{|\BM(t)-f(t)|}{\phi(|t|)}=0\quad \textrm{a.s.}\, ,\ee
which holds thanks to the property \eqref{Bphitozero} and the boundedness assumption on $f$ [i.e., $f(t)/\phi(|t|)\rightarrow 0$].
Equation \eqref{largeTlim} follows.

The limit \eqref{largeTlim} implies the existence of some value $T_\star>0$ such that 
\be\label{PaboveTstar} \prob(\mathscr E_{f,\phi,T}) >0 \quad \textrm{for all $T\geq T_\star$}\, . \ee
It thus remains to prove the theorem for $0<T<T_\star$. 

To that aim let us split $\mathscr E_{f,\phi,T}$ at $|t|=T_\star$ and identify it as a joint event $\mathscr U_{f,\phi,T,T_\star}\bigcap \mathscr E_{f,\phi,T_\star}$ where
%
%
%
\begin{equation}\label{fromTtoTstar}\begin{split} \mathscr U_{f,\phi,T,T_\star} = \big\{ |\BM(t)-f(t)|<\phi(|t|)\, , & \\  \forall\, T\leq |t| \leq T_\star & \big\}\, . \end{split}\end{equation}
Our strategy is to show first that these two events are generic, and then apply the techniques of the local extended forgery (see Section \ref{extendedforgeryproof}) to derive the genericity of their intersection. 
The case of $\mathscr E_{f,\phi,T_\star}$ is taken care of by Eq.~\eqref{PaboveTstar}.
For \eqref{fromTtoTstar} observe that $\mathscr U_{f,\phi,T,T_\star}\supset\mathscr U_{f,\delta,T_*}$ whenever $\delta$ is chosen below the minimum value of the continuous function $\phi(|t|)$ over the compact time domain $T\leq|t|\leq T_*$, since $|\BM(t)-f(t)|<\delta \leq \phi(|t|)$.
If we pick a $\delta>0$ in this way (this is possible since $\phi$ is positive), we then obtain $\prob(\mathscr U_{f,\phi,T,T_\star})\geq \prob(\mathscr U_{f,\delta,T_*})>0$ by monotonicity of $\prob(\cdot)$ and Levy's forgery theorem. 
The last step is to ensure that the intersection is generic too.
This is analogous to the extended forgery theorem \eqref{fullextendedforgery} and can be proven along the exact same lines.
\qed

\medskip\noindent\textbfsf{Consequences for the other forgery theorems.} 
Strong versions of the subsequent forgery theorems can be obtained by replacing the event \eqref{linearapprox} with its generalization \eqref{nonlinearapprox}.
The end result is to weaken the convergence assumptions required for the random variables $\X,\Y$ and therefore widen the applicability of the forgery of statistical dependences and the Brownian independence test.
We gather the results here without repeating the derivations.

The strong extended forgery theorem states that {\it if $f$ is bounded and continuous with $f(0)=0$, $\phi$ is a growth function as above, and $\delta,T>0$, then}
\be \prob(\mathscr U_{f,\delta,T} \textstyle\bigcap \mathscr E_{f,\phi,T}) >0\, . \ee
A covariance error inequality similar to Eq.~\eqref{coverrorbound} then holds for arbitrary sample paths $\BM\in\mathscr U_{f,\delta,T}\bigcap\mathscr E_{f,\phi,T}$ and $\BM'\in\mathscr U_{g,\delta,T}\bigcap\mathscr E_{g,\phi,T}$  with the bound
%
%
%
\be\label{strongbound} \begin{split} \varepsilon_{\textsc b}(\delta,\phi) = \ & 2(M_f+M_g) \delta +2\delta^2\\ &+2\big[M_g\langle\phi(|\X|)\rangle + M_f\langle\phi(|\Y|)\rangle\big] \\ &+2\langle\phi(|\X|)+\phi(|\Y|)\rangle \delta \\ &+\langle\phi(|\X|)\phi(|\Y|)\rangle+\langle\phi(|\X|)\rangle\langle\phi(|\Y|)\rangle \end{split}\ee
instead of \eqref{bound}. 
As a consequence, the forgery theorem of statistical dependences and the Brownian independence test both hold under the weaker convergence conditions that {\it there exists a growth function $\phi$ for which $\langle\phi(|\X|)\rangle$, $\langle\phi(|\Y|)\rangle$, and $\langle\phi(|\X|)\phi(|\Y|)\rangle$ are finite.}

\section{MEASURABILITY OF STOCHASTIC COVARIANCE}\label{measurabilityproof}
The premise of all the forgery theorems is that the associated approximation events are measurable. 
For the sets \eqref{unifapprox}, \eqref{linearapprox} enforcing algebraic conditions on a time domain, this is a general property of separable stochastic processes such as Brownian motion \cite{GikhmanSkorokhod1969book}. Indeed, the principles of probability theory only allow constraints on at most a countable infinity of time points but the notion of separability enables considering continuous time domains as well. In a nutshell, Brownian motion is separable because conditions such as $a\leq \BM(t)\leq b$ imposed on a dense grid of rational time points $t=k/n$ extend automatically to nonrational times by sample path continuity.

The story is different for the set \eqref{covapprox} because it involves a functional 
\be\label{covBB} \cov[\BM(\X),\BM'(\Y)] = \langle\BM(\X)\BM'(\Y)\rangle-\langle\BM(\X)\rangle\langle\BM'(\Y)\rangle\ee 
of the sample paths $\BM,\BM'$ that depends in particular on the distribution of the random times $\X,\Y$.
Here we provide convergence conditions that ensure the measurability of this functional and thus of the covariance approximation event \eqref{covapprox}.

\medskip\noindent\textbfsf{Sufficient conditions for measurability.} {\it The expectation values $\langle\BM(\X)\rangle$, $\langle\BM'(\Y)\rangle$, and $\langle\BM(\X)\BM'(\Y)\rangle$ are measurable functionals of the sample paths $\BM,\BM'$ whenever $\langle |\X|^{1/2} \rangle$, $\langle |\Y|^{1/2} \rangle$, and $\langle |\X\Y|^{1/2} \rangle$ are finite, respectively. When these three conditions hold, the stochastic covariance \eqref{covBB} is thus measurable too.}

It is noteworthy that these conditions can be replaced by the stronger constraint that $\X,\Y$ are $L_1$, i.e., $\langle|\X|\rangle$ and $\langle|\Y|\rangle$ are finite (since the Cauchy-Schwarz inequality implies, e.g., $\langle |\X\Y|^{1/2} \rangle \leq \langle |\X| \rangle^{1/2}\langle |\Y| \rangle^{1/2}$).
In Section \ref{statforgery} we used the even stronger assumption that $\X,\Y$ are $L_2$ to simplify and shorten our formulation of results.

\medskip\noindent\textbfsf{Proof.} This is a direct consequence of the Fubini-Tonelli theorem \cite{GikhmanSkorokhod1969book}. 
Let us focus on the functional $\BM\rightarrow\langle \BM(\X)\rangle$ [the argument for $\langle \BM'(\Y)\rangle$ and $\langle \BM(\X)\BM'(\Y)\rangle$ is completely similar].
In this setup, the theorem ensures its measurability if the iterated expectation value of $\BM(\X)$, computed first by averaging over sample paths $\BM$ and then over random times $\X$, is absolutely convergent. 
Since $\BM(t)$ is a Gaussian random variable with zero mean and variance $|t|$, we find explicitly
\be\label{FBcondition1} \big\langle \langle | \BM(t) |\rangle_{\textrm{$t=\X$}} \big\rangle = k\,  \big\langle |\X|^{1/2}\big\rangle\, ,\ee
where 
\be k=\frac{1}{\sqrt{2\pi}}\int_{-\infty}^\infty \mathrm{d} z\, |z|\, \mathrm{e}^{-z^2/2}<\infty\, .\ee
Thus the finiteness condition on $\langle |X|^{1/2}\rangle$ indeed fulfills the requirement of the Fubini-Tonelli theorem. 

Technically this application also needs the extra prerequisite that $\BM(\X)$ be a random variable, i.e., that the evaluation map $(\BM,\X)\rightarrow \BM(\X)$ be \emph{jointly} measurable (and not just at fixed path $\BM$ or fixed time $\X$).
This is a property of continuous stochastic processes such as Brownian motion \cite{GikhmanSkorokhod1969book}.
\qed

%
\section{COMPLETION OF PROOFS} \label{technicalappendix}
\renewcommand\theequation{\Alph{section}.\arabic{equation}}
\setcounter{equation}{0}

\subsection{Markovian decomposition formula \eqref{markovintegral}}\label{technicalappendix-markov}

This equation relies on the weak Markov property and is actually valid for general Markovian processes.
Denoting by $\mathfrak U_{f,\delta,T}$ the $\sigma$-algebra generated by the event $\mathscr U_{f,\delta,T}^+$, we can write 
\be\label{condtoG} \prob(\mathscr U_{f,\delta,T}^+\ {\textstyle\bigcap}\ \mathscr E_{f,v,T}^+) = \big\langle \boldsymbol 1_{\mathscr U_{f,\delta,T}^+}\, \prob[\mathscr E_{f,v,T}^+ \, | \, \mathfrak U_{f,\delta,T}] \big\rangle \ee
merely by definition of the conditional probability $\prob[\mathscr E_{f,v,T}^+ \, | \, \mathfrak U_{f,\delta,T}]$  \cite{GikhmanSkorokhod1969book}.
Furthermore, $\mathfrak U_{f,\delta,T}$ is contained in the $\sigma$-algebra $\mathfrak F_T$ generated by the process $\BM(t)$ with $0\leq t\leq T$, since $\mathscr U_{f,\delta,T}^+$ only involves conditions on this time interval.
From this observation,
\begin{align} \nonumber \prob[\mathscr E_{f,v,T}^+ \, | \, \mathfrak U_{f,\delta,T}] &= \big\langle \prob[\mathscr E_{f,v,T}^+ \, | \, \mathfrak F_T] \ \big | \ \mathfrak U_{f,\delta,T} \big\rangle \\ \label{markov} &= \big\langle \prob[\mathscr E_{f,v,T}^+ \, | \, \BM(T)] \ \big | \ \mathfrak U_{f,\delta,T} \big\rangle\, , \end{align}
where $\langle\, \cvd \, | \, \mathfrak U_{f,\delta,T}\rangle$ denotes the expectation value conditional to $\mathfrak U_{f,\delta,T}$.
In the first equality we applied the tower property of conditioning that follows from $\mathfrak U_{f,\delta,T}\subset\mathfrak F_T$. In the second we used the weak Markov property that events involving conditions for $t\geq T$ (such as $\mathscr E_{f,v,T}^+$) only depend on their past via the ($\sigma$-algebra generated by the) random variable $\BM(T)$.

Combining Eqs.~\eqref{condtoG} and \eqref{markov}, we therefore find that the left-hand side of Eq.~\eqref{markovintegral} is equal to 
$$\big\langle \boldsymbol 1_{\mathscr U_{f,\delta,T}^+}\, \langle\prob[\mathscr E_{f,v,T}^+ \, | \, \BM(T)] \, \big| \, \mathfrak U_{f,\delta,T}\rangle \big\rangle\, .$$
The conditional expectation to $\mathfrak U_{f,\delta,T}$ can be dropped in this expression, again by mere definition of conditioning to $\mathfrak U_{f,\delta,T}$, so we recover the right-hand side of \eqref{markovintegral}.

\subsection{Representation \eqref{condtoBT} of $\boldsymbol{\prob[\mathscr E_{f,v,T}^+ \, | \, \BM(T)]}$}\label{technicalappendix-condtoBT}

The defining property of the conditional probability $\prob[\mathscr E_{f,v,T}^+ \, | \, \BM(T)]$ is that
%
%
\be\label{conditioning}\begin{split} \prob\big(\{\BM(T) & \in S\}  \textstyle\bigcap \mathscr E_{f,v,T}^+\big) \\&= \big\langle \boldsymbol 1_{\{\BM(T)\in S\}} \, \prob[\mathscr E_{f,v,T}^+ \, | \, \BM(T)] \big\rangle\end{split}\ee
for all Borel sets $S\subset \mathbb R$, which only characterizes it modulo a zero-probability set.
To prove the statement \eqref{condtoBT}, it thus suffices to check that $p_{f,\delta,T}(\BM(T))$ satisfies the same property.
This is a consequence of the fact that Brownian motion has independent increments themselves distributed as Brownian motions \cite{GikhmanSkorokhod1969book}. 

Indeed, the left-hand side of Eq.~\eqref{conditioning} is computed as the integral 
over all  sample paths $\BM$ satisfying both $\BM(T)\in S$ and $|\BM(t)-f(t)|<v t$ for all $t\geq T$. 
Defining the increment $\BM'(t')=\BM(T+t')-\BM(T)$ with time step $t'=t-T$, the latter constraint becomes 
\be |\BM'(t')+\BM(T)-f(T+t')|<v (T+t')=\delta+vt' \ee
for all $t'\geq 0$, or equivalently $\BM' \in \mathscr A_{f,\delta,T}^{\BM(T)}$ in view of the definition \eqref{defAx}.
Now as recalled above $\BM'$ is a Brownian motion independent of the position $\BM(T)$. 
An application of the Fubini-Tonelli theorem \cite{GikhmanSkorokhod1969book} thus enables us to integrate first over all $\BM' \in \mathscr A_{f,\delta,T}^{\BM(T)}$ at fixed $\BM(T)$ and then over $\BM(T)\in S$.
Therefore we obtain
%
%
%
\begin{align} \nonumber \prob\big(\{\BM(T) & \in S\} \textstyle\bigcap \mathscr E_{f,v,T}^+\big) \\ \nonumber &= \big\langle \boldsymbol 1_{\{\BM(T)\in S\}} \, \langle \boldsymbol 1_{\{\BM'\in \mathscr A_{f,\delta,T}^{x}\}} \rangle_{x=\BM(T)} \big\rangle \\ \label{tutu} &= \big\langle \boldsymbol 1_{\{\BM(T)\in S\}} \,p_{f,\delta,T}\big(\BM(T)\big) \big\rangle\, , \end{align}
where we used the identity $\big\langle \boldsymbol 1_{\{\BM'\in \mathscr A_{f,\delta,T}^{x}\}} \big\rangle=\prob(\mathscr A_{f,\delta,T}^{x})$ and the definition \eqref{psidef} in passing through the second equality.
This ends the demonstration of Eq.~\eqref{condtoBT}.

\subsection{Continuity property \eqref{Pcontinuous} of $\boldsymbol{p_{f,\delta,T}}$}\label{technicalappendix-continuity}

Let us consider an arbitrary sequence $x_n$ converging to some $x$ inside the bottleneck interval at $t=T$.
We shall then prove that the limit of $p_{f,\delta,T}(x_n)=\prob(\mathscr A_{f,\delta,T}^{x_n})$ as $n\rightarrow\infty$ exists and equals to $p_{f,\delta,T}(x)=\prob(\mathscr A_{f,\delta,T}^{x})$. 

We start from the difference formula
%
%
%
\be\label{Pdiff}\begin{split} \prob(\mathscr A_{f,\delta,T}^{x}) - \prob(\mathscr A_{f,\delta,T}^{x_n}) = \ &  \prob(\mathscr A_{f,\delta,T}^{x} {\setminus} \mathscr A_{f,\delta,T}^{x_n}) \\  & - \prob(\mathscr A_{f,\delta,T}^{x_n} {\setminus} \mathscr A_{f,\delta,T}^{x}) \end{split}\ee
and argue that each of the two terms in the right-hand side converge to zero. Since sequences of probabilities are nonnegative, it is actually sufficient to show that   
\be\label{limsup1} \limsup_{n\rightarrow\infty} \prob(\mathscr A_{f,\delta,T}^{x} {\setminus} \mathscr A_{f,\delta,T}^{x_n}) = \prob(\mathscr A_{f,\delta,T}^{x} {\setminus} \mathscr A_{f,\delta,T}^{x_n} \ \textrm{i.o.}) \ee 
and
\be \label{limsup2} \limsup_{n\rightarrow\infty} \prob(\mathscr A_{f,\delta,T}^{x_n} {\setminus} \mathscr A_{f,\delta,T}^{x}) = \prob(\mathscr A_{f,\delta,T}^{x_n} {\setminus} \mathscr A_{f,\delta,T}^{x} \ \textrm{i.o.})\ee
%
%
vanish.
Here  i.o.~stands for ``infinitely often'' and means that an infinite number of events in the sequence occur, and indeed allows evaluating superior limits \cite{GikhmanSkorokhod1969book}. 

To proceed with the proof, let us introduce the \emph{zero-probability} set $\mathscr Z$ of paths that are not continuous or do not satisfy the law of large numbers $\lim_{t\rightarrow\infty}\BM(t)/t=0$, and the \emph{boundary-hitting} event
\be\label{hitevent} \begin{split}\mathscr B_{f,\delta,T}^{x} = \big\{  & |\BM(t)+x-f(T+t)|\leq\delta+vt \\  & \textrm{$\forall\, t\geq 0$, with equality at some $t=\tau$}\big\}\end{split}\ee
that is closely related to \eqref{defAx} but imposes that sample paths actually reach the boundary of the associated neighborhood.
This event turns out to also have probability zero, because a Brownian path hitting the boundary must a.s.~also cross it and thus leave the neighborhood. 
For completeness we provide a derivation of this boundary-crossing law in \ref{crossinglaw}.

With these definitions at hand, we claim that 
\begin{align}\label{incl1} \{\mathscr A_{f,\delta,T}^{x} {\setminus} \mathscr A_{f,\delta,T}^{x_n} \ \textrm{i.o.}\}  & \subset \mathscr Z\, , \\ \label{incl2} \{\mathscr A_{f,\delta,T}^{x_n} {\setminus} \mathscr A_{f,\delta,T}^{x} \ \textrm{i.o.}\} & \subset \mathscr Z \, \textstyle\bigcup\, \mathscr B_{f,\delta,T}^{x}\, .\end{align}
They imply directly that the two limits \eqref{limsup1} and \eqref{limsup2} vanish by monotonicity and subadditivity of $\prob(\cdot)$, because $\prob(\mathscr Z)=\prob(\mathscr B_{f,\delta,T}^{x}) = 0$.
Therefore the right-hand side of Eq.~\eqref{Pdiff} does indeed tend to zero as $n\rightarrow\infty$.

It remains to derive these two inclusions.

\medskip\noindent{\textbfsf{First inclusion \eqref{incl1}.}} 
Let us consider an arbitrary sample path $\BM$ satisfying $\BM\in\mathscr A_{f,\delta,T}^{x} {\setminus} \mathscr A_{f,\delta,T}^{x_n}$ i.o. 
Explicitly, this means that we can build a diverging sequence $n_k$ of indices such that $\BM\in\mathscr A_{f,\delta,T}^{x}$ and $\BM\notin\mathscr A_{f,\delta,T}^{x_{n_k}}$, i.e., 
\be\label{io1} |\BM(t_k)+x_{n_k}-f(T+t_k)| \geq \delta+vt_k\ee
for some $t_k\geq 0$. At this point two cases emerge. Both will lead to the conclusion that $\BM\in\mathscr Z$, which corresponds to the sought inclusion \eqref{incl1}.

\emph{(i) The set of times $t_k$ is bounded.} 
In that situation, the Bolzano-Weierstrass theorem provides a subsequence $t_{k_m}$ converging to some finite time $\tau$. 
Assuming that $\BM$ is continuous, let us evaluate Eq.~\eqref{io1} along $k=k_m$ and take $m\rightarrow\infty$. 
Using $x_{n_{k_m}}\rightarrow x$ and the continuity of $\BM$ and $f$, we find $|\BM(\tau)+x-f(T+\tau)|\geq\delta+v\tau$, which contradicts the fact that $\BM\in\mathscr A_{f,\delta,T}^{x}$. 
Therefore $\BM$ cannot be continuous at $\tau$, henceforth $\BM\in\mathscr Z$.

\emph{(ii) The set of times $t_k$ is unbounded.}
Then we select a diverging subsequence $t_{k_m}$, divide both sides of Eq.~\eqref{io1} by $t_{k}$, and set $k=k_m$ with $m\rightarrow\infty$. 
Since $t_{k_m}\rightarrow\infty$ and both $x_{n}$ and $f$ are bounded, it follows that $\limsup_{m\rightarrow\infty} |\BM(t_{k_m})|/t_{k_m} \geq v >0$.
This shows that the sample path $\BM$ cannot satisfy the law of large numbers, hence $\BM\in\mathscr Z$.

\medskip\noindent{\textbfsf{Second inclusion \eqref{incl2}.}} Similarly, let us now consider a sample path satisfying $\BM\in\mathscr A_{f,\delta,T}^{x_n} {\setminus} \mathscr A_{f,\delta,T}^{x}$ i.o., meaning that there exists a diverging sequence $n_k$ such that
%
%
%
\be\label{io2}|\BM(t)+x_{n_k}-f(T+t)| < \delta+vt \ee
for all $t\geq 0$, and a set of times $t_k\geq 0$ such that
\be\label{io3} |\BM(t_k)+x-f(T+t_k)| \geq \delta+vt_k\, . \ee
Taking $k\rightarrow\infty$ in Eq.~\eqref{io2} yields the nonstrict inequality $|\BM(t)+x-f(T+t)|\leq\delta+vt$.
Repeating our analysis of Eq.~\eqref{io1} to Eq.~\eqref{io3} also shows that equality must hold at some time $\tau$, i.e., $\BM\in\mathscr B_{f,\delta,T}^{x}$, or else $\BM\in\mathscr Z$, so we obtain the sought inclusion \eqref{incl2}.

\subsection{The boundary-crossing law for $\boldsymbol{\prob(\mathscr B_{f,\delta,T}^x)=0}$}\label{crossinglaw}
It is convenient to introduce the upper ($+$ sign) and lower ($-$ sign) boundary curves
\be\label{boundarycurves} \gamma_{\pm} (t) = f(T+t) - x \pm (\delta+vt)\, . \ee
They inherit the continuity of $f$, and since $x$ lies within the bottleneck interval at $t=T$ they also satisfy the condition $\gamma_-(0)<0<\gamma_+(0)$.

The boundary-crossing law that we shall derive can be stated formally as follows:
\begin{align}\label{maxrule} \prob \big( \max_{t\geq 0} [\BM(t)-\gamma_+(t)] = 0 \big) & = 0\, , \\ \label{minrule}\prob \big( \min_{t\geq 0} [\BM(t)-\gamma_-(t)] = 0 \big) & = 0\, .\end{align}
This provides the sought estimate of $\prob(\mathscr B_{f,\delta,T}^x)$. 
Indeed, observing that the boundary-hitting event \eqref{hitevent} implies either $\BM(t)\leq\gamma_+(t)$ with equality at some time, or else $\BM(t)\geq\gamma_-(t)$ with equality at some time, we find
%
%
%
\be\label{minmax}\begin{split} \mathscr B_{f,\delta,T}^x \subset  \big\{ & \max_{t\geq 0} [\BM(t)-\gamma_+(t)] = 0 \big\} \\ & {\textstyle\bigcup}\ \big\{ \min_{t\geq 0} [\BM(t)-\gamma_-(t)] = 0 \big\}\, . \end{split}\ee
%
Applying the monotonicity and subadditivity of $\prob(\cdot)$ and the boundary-crossing law then ends the derivation of $\prob(\mathscr B_{f,\delta,T}^x)=0$.

\medskip\noindent{\textbfsf{Proof of the boundary-crossing law.}} 
We shall focus here on the derivation of Eq.~\eqref{maxrule} [the case for \eqref{minrule} is completely similar].
To simplify the notations, let us introduce the running maximum 
\be\mathsf M_{\tau} = \max_{0\leq t\leq\tau}[\BM(t)-\gamma_+(t)] \quad \textrm{for $0<\tau\leq\infty$} \ee
(with the proviso that, say, $\mathsf M_\tau=-\infty$ if a maximum is not attained for $0\leq t\leq\tau$).
Thus Eq.~\eqref{maxrule} simply reads $\prob(\mathsf M_{\infty} = 0)=0$.
In what follows we fix $\tau<\infty$ and we will eventually let $\tau\rightarrow0$. 

The hitting (without crossing) event $\mathsf M_\infty = 0$ implies that either $\BM(t)$ reaches $\gamma_+(t)$ before $t=\tau$, which corresponds to the event $\mathsf M_\tau=0$, or else the increment $\BM'(t')=\BM(\tau+t')-\BM(\tau)$ reaches $\gamma_+(\tau+t')-\BM(\tau)$ for $t'=t-\tau>0$, which corresponds to the hitting event $\mathsf W=-\BM(\tau)$, where we introduced the random variable
\be \mathsf W = {\textstyle \max}_{t >0}[\BM'(t)-\gamma_+(\tau+t)]\, . \ee
It thus follows that
\be\label{M0bound}\prob(\mathsf M_{\infty}=0) \leq \prob(\mathsf M_{\tau}=0) + \prob\big[\mathsf W=-\BM(\tau)\big]\ee
by monotonicity and subadditivity of $\prob(\cdot)$. 
To obtain Eq~\eqref{maxrule}, we now evaluate the two terms in the right-hand side and show that both vanish as $\tau\rightarrow0$.

For the first term it is sufficient to show that
\be\label{M0bound-firstterm} \limsup_{\tau\rightarrow0} \prob(\mathsf M_{\tau}=0)=\prob(\mathsf M_{\tau}=0\ \textrm{i.o.})=0\, .\ee
The limit event $\mathsf M_{\tau}=0$ i.o.~means explicitly that $\BM(t)$ hits $\gamma_+(t)$ for arbitrarily small times $t>0$ [the case $t=0$ being ruled out by $\BM(0)=0<\gamma_+(0)$].
This implies that $\lim_{t\rightarrow0} \BM(t) = \gamma_+(0)>\BM(0)$ so $\BM$ cannot be continuous at $t=0$ and the probability that $\mathsf M_\tau=0$ occurs i.o.~is thus zero.

The second term vanishes for all $0<\tau<\infty$.
Noting that the increment process $\BM'$ (and thus, $\mathsf W$) is independent of $\BM(\tau)$ and applying the Fubini-Tonelli theorem, we can write explicitly
\be\label{M0bound-secondterm} \prob\big[\mathsf W=-\BM(\tau)\big] = \int_{-\infty}^{\infty} \mathrm{d} w \, \frac{\exp(-w^2/2\tau)}{\sqrt{2\pi\tau}} \ \prob(\mathsf W=-w)\,. \ee
But it is a general property of probability theory that $\prob(\mathsf W=-w)$ may be nonzero for at most a countable set of values $w$. 
Therefore $\prob(\mathsf W=-w)=0$ for almost all $w$ and thus the integral \eqref{M0bound-secondterm} vanishes.


%

\subsection{Bound \eqref{coverrorbound} for the covariance error}\label{technicalappendix-bound}

Let us introduce the shorthand notations 
%
%
%
\be \begin{aligned} &\Delta \cov=\cov[\BM(\X),\BM'(\Y)]-\cov[f(\X),g(\Y)]\, , \\ &\Delta \BM=\BM(\X)-f(\X)\, , \quad \Delta \BM'=\BM'(\Y)-g(\Y)\, . \end{aligned} \ee
Using the bilinearity of covariance and the triangle inequality, we have 
%
%
%
\be\label{deltacov} \begin{split} \big|\Delta\cov \big| \leq \big|\cov[\Delta\BM&,g(\Y)]\big| + \big|\cov[f(\X),\Delta\BM']\big| \\ &+ \big|\cov(\Delta\BM,\Delta\BM')\big|\, .\end{split} \ee
We start by examining the first term.
Developing the covariance, applying triangle inequalities, and using the boundedness of $g$, we find
\begin{align} \big|\cov[\Delta\BM,g(\Y)]\big| &= \big| \langle\Delta\BM\, g(\Y)\rangle -\langle\Delta\BM\rangle\langle g(\Y)\rangle\big| \nonumber \\ &\leq \langle |\Delta\BM\, g(\Y)|\rangle+\langle|\Delta\BM|\rangle\langle |g(\Y)|\rangle \nonumber \\ \label{}&\leq 2 M_g \langle |\Delta\BM |\rangle \nonumber \\ \label{covDeltaBg} &< 2 M_g \big(\delta+v \langle|\X|\rangle\big)\, . \end{align}
The last inequality was obtained by splitting the expectation value $\langle|\Delta\BM|\rangle$ into its contributions from the two regions
\renewcommand{\labelitemi}{$\boldsymbol \cdot$}
\begin{itemize}
\itemsep -0.3em
\item $\{|\X|\leq T\}$ where $|\Delta\BM|< \delta$ [see Eq.\eqref{unifapprox}], and
\item $\{|\X|> T\}$ where $|\Delta\BM|< v |\X|$ [see Eq.~\eqref{linearapprox}].
\end{itemize}
The second term is completely analogous,
\be\label{covfDeltaB} \big|\cov[f(\X),\Delta\BM']\big| < 2 M_f \big(\delta+v \langle|\Y|\rangle\big)\, . \ee
For the last term, we use four regions: 
\begin{itemize}
\itemsep -0.3em
\item $\{|\X|\leq T, |\Y|\leq T\}$ where $|\Delta\BM\, \Delta\BM' |<\delta^2$, 
\item $\{|\X|\leq T, |\Y|> T\}$ where $|\Delta\BM\, \Delta\BM' |<\delta v |\Y|$, 
\item $\{|\X|> T, |\Y|\leq T\}$ where $|\Delta\BM\, \Delta\BM' |<\delta v |\X|$, and
\item $\{|\X|> T, |\Y|> T\}$ where $|\Delta\BM\, \Delta\BM' |<v^2 |\X\Y|$. 
\end{itemize}
We then obtain in a similar way
%
%
%
\begin{align} \big|\cov(\Delta\BM,\Delta\BM')\big| &\leq \langle|\Delta\BM\, \Delta\BM'|\rangle+\langle|\Delta\BM|\rangle\langle |\Delta\BM'|\rangle \nonumber \\ \label{covDeltaBDeltaB} \begin{split} & < 2\delta^2 + 2\delta v \langle|\X|\rangle+ 2\delta v \langle|\Y|\rangle \\ & \quad +
v^2 \big(\langle|\X\Y|\rangle+\langle|\X|\rangle\langle|\Y|\rangle\big)\, . \end{split} \end{align}
The covariance error inequality \eqref{coverrorbound} with bound \eqref{bound} is recovered by combining Eqs.~\eqref{deltacov} to \eqref{covDeltaBDeltaB}. 



\bibliography{./references}

\end{document}